\documentclass[prb,twocolumn,showpacs,preprintnumbers]{revtex4}

\usepackage{graphicx}
\usepackage{dcolumn}
\usepackage{amsmath}
\usepackage{color}

\begin{document}

\title{Metal-insulator transition and local-moment collapse in FeO under
pressure}

\author{I. Leonov}
\affiliation{Theoretical Physics III, Center for Electronic Correlations and
Magnetism, Institute of Physics, University of Augsburg, 86135 Augsburg, Germany}

\date{\today}

\begin{abstract}
We employ a combination of the \emph{ab initio} band structure methods and 
dynamical mean-field theory to determine the electronic structure and phase 
stability of paramagnetic FeO at high pressure and temperature. Our results 
reveal a high-spin to low-spin transition within the B1 crystal structure of 
FeO upon compression of the lattice volume above 73~GPa. The spin-state 
transition is accompanied by an 
orbital-selective Mott metal-insulator transition (MIT). The lattice volume 
is found to collapse by about 8.5~\% at the MIT, implying a complex interplay 
between electronic and lattice degrees of freedom. Our results for the electronic 
structure and lattice properties are in overall good agreement with experimental 
data.

\end{abstract}

 \maketitle

\section{Introduction}

Iron monoxide, w\"ustite (FeO), is a basic oxide component of the Earth's interior.
Its electronic state and phase stability is of fundamental importance 
for understanding the properties and evolution of the Earth's lower mantle 
and outer core. \cite{Phase-diagram}
Nevertheless, in spite of long-term intensive research, the phase diagram of
FeO at high pressures and temperatures, as well as several key properties, such 
as electronic structure and magnetic properties, are still poorly understood.

FeO has a relatively complex pressure-temperature phase diagram with at least
five allotrops. \cite{Phase-diagram} Under ambient conditions, it is a paramagnetic 
Mott insulator with a rock-salt B1 crystal structure. Upon compression above 
$\sim 16$~GPa, B1-type FeO undergoes a structural transition to the rhombohedral 
$R\bar{3}$ phase (rB1), \cite{rB1-phase} which further transforms to the NiAs (B8) 
structure above 90~GPa. \cite{B8-phase}
Under ambient pressure, FeO is antiferromagnetic below the N\'eel 
temperature $T_N \sim 198$~K. \cite{T-Neel}
The high-pressure properties of FeO have attracted much recent interest 
both from a theoretical and experimental point of views. 
\cite{IC93,MA97,MF98,ZT98,GC03,TB06,KM08,SP10,OC12,HS15}
Shock-wave compression and electrical conductivity experiments reported a
possible existence of a high-pressure metallic phase of FeO above $\sim
70$~GPa. \cite{KJ86} 
On the basis of high-pressure M\"ossbauer spectroscopy measurements, \cite{PT97}
the metallic state was assigned to a high-spin (HS) to low-spin (LS) transition,
which has been proposed by first-principles calculations to occur above $\sim$
100-200~GPa. \cite{IC93}
In contrast to that, x-ray emission spectroscopy indicates that the Fe high-spin
state is preserved at least up to $\sim 140$~GPa (at room temperature), \cite{BS99}
while collapsing to the LS state upon further heating. \cite{MR07}
On the basis of these measurements, the insulator-to-metal HS-LS transition has
long been considered to be due to a structural transformation from the B1 to B8
lattice.
Nevertheless, recent experiments have shown that the B1-type FeO undergoes a
high-temperature insulator-to-metal transition at about 70~GPa, retaining the
same lattice structure. \cite{OC12} The observed metallic phase was proposed to
relate to a spin-state crossover.
Moreover, it has been shown that the B1-type structure remains stable at high
pressure and temperature, being the stable phase along the geotherm through the
Earth's mantle and outer core. \cite{Phase-diagram, FC10}

These recent experiments have lead us to reinvestigate the electronic structure
and local magnetic state of Fe in FeO at high pressure and temperature by employing a combination of
the \emph{ab initio} electronic structure methods and dynamical mean-field
theory of strongly correlated electrons (LDA+DMFT). \cite{DMFT,LDA+DMFT} (LDA
stands here for the local density approximation).
Applications of LDA+DMFT have shown to capture all generic aspects of a Mott
metal-insulator transition, such as a coherent quasiparticle behavior, formation
of the lower- and upper-Hubbard bands, and strong renormalization of the
effective electron mass, providing a good quantitative description of the
electronic and lattice  properties. In particular, we employ an important
advance of the LDA+DMFT approach which is able to determine the electronic 
structure and phase stability of correlated materials.
\cite{SK01,HK04,HM01,SH06,AB06,KA07,LB08,KL08,DM09,LP11,AP11,HW12,GP12,PMM14,PM14,LA14}
We use this advanced theory to investigate the electronic structure and phase
stability of paramagnetic FeO at high pressure and temperature, which remained unexplored up to now.
We find that magnetic collapse occurs in the B1 crystal structure of
paramagnetic FeO upon compression to $\sim 73$~GPa, in agreement with
experiment. The HS-LS transition is intimately linked with an orbital-selective
Mott metal-insulator transition, which is accompanied with a collapse of the
lattice volume by about 8.5~\% at the MIT.
Our results for the electronic structure and lattice properties are in overall 
good agreement with experimental data.

\section{Method}

In this work, we investigate the electronic structure and phase stability of paramagnetic FeO under pressure using the GGA+DMFT computational 
approach (GGA: generalized gradient approximation). To this end, we calculate the total energy and local moments of the B1 cubic crystal structure of FeO as a function of
lattice volume. Below we denote the compressed phase by the relative volume
w.r.t. the calculated equilibrium lattice volume as $\nu \equiv V/V_0$.
We employ a fully charge self-consistent GGA+DMFT approach \cite{charge-sc-LDA+DMFT}
implemented with plane-wave pseudopotentials \cite{Pseudo}. For the partially 
filled Fe $3d$ and O $2p$ orbitals we construct a basis set of atomic-centered 
symmetry-constrained Wannier functions \cite{MV97,Wannier-functions}. To solve the 
realistic many-body problem, we employ the continuous-time hybridization-expansion 
quantum Monte-Carlo algorithm \cite{ctqmc}. The calculations are performed in the 
paramagnetic state at temperature $T = 1160$~K. We use the average Coulomb 
interaction $U=7$~eV and Hund's exchange $J=0.89$~eV for the Fe $3d$ shell, 
in accordance with previous estimates \cite{GC03,MA97,SP10,OC12}. The Coulomb 
interaction is treated in the density-density approximation. The spin-orbit 
coupling is neglected in these calculations. 
The $U$ and $J$ values are assumed to remain constant upon variation of the
lattice volume. In addition, we check how our results depend on a possible
reduction of the $U$ value under pressure. For this purpose, we perform
calculations with a substantially smaller value, $U=5$~eV. We employ the 
fully-localized double-counting correction, evaluated from the self-consistently 
determined local occupancies, to account for the electronic interactions already 
described by GGA. The spectral functions were computed using the maximum entropy 
method. The angle resolved spectra were evaluated from analytic continuation of 
the self-energy using Pad\'e approximants.

\begin{figure}[tbp!]
\centerline{\includegraphics[width=0.5\textwidth,clip=true]{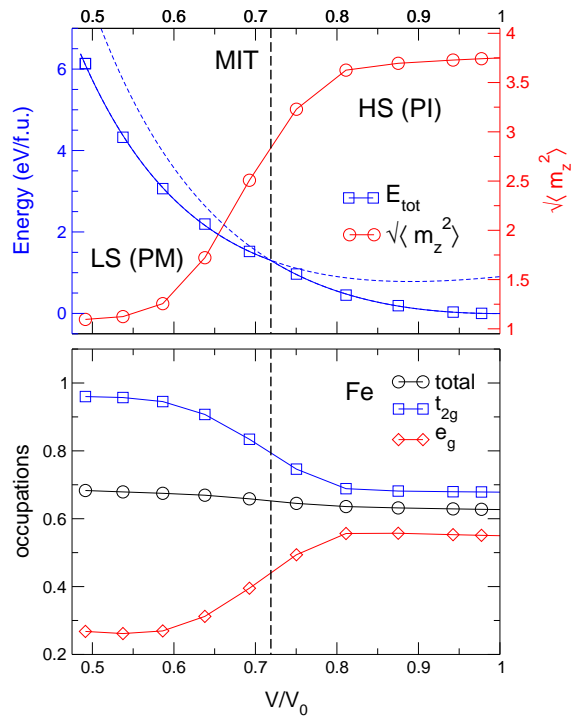}}
\caption{Top: Total energy and local moment $\sqrt{\langle m_z^2
\rangle}$ of paramagnetic FeO calculated by GGA+DMFT as a function of
lattice volume. The lattice collapse associated with the HS-LS state transition
is depicted by a vertical black dashed line. Bottom: Fe $3d$ and partial
$t_{2g}$/$e_g$ occupations as a function of volume.
}
\label{Fig_1}
\end{figure}

\section{Results and discussion}

We first compute the electronic structure of FeO within non-spin-polarized
GGA using the plane-wave pseudopotential approach. \cite{Pseudo} Overall, our 
results agree well with previous band-structure calculations. In particular, the 
calculated equilibrium lattice constant is found to be $a=7.74$~a.u. The calculated 
bulk modulus is $B=232$~GPa. 
We note however that GGA calculations give a metallic solution, in qualitative
disagreement with experiment. 
The calculated equilibrium lattice constant is remarkably smaller, by $\sim$ 6-7~\%, 
than the experimental one. We also note a strong overestimation of the
bulk modulus, which is more than $\sim 25$~\% off the experimetal value.
Clearly, standard band-structure techniques cannot explain the properties of
paramagnetic FeO, since they do not treat electronic correlations adequately.


To resolve this obstacle, we now compute the electronic structure and phase
stability of FeO using the fully charge self-consistent GGA+DMFT method. In
Fig.~\ref{Fig_1} (top) we show the evolution of the total energy and local
magnetic moment $\sqrt{\langle m_z^2 \rangle}$ of paramagnetic FeO as a function
of lattice volume. 
We fit the calculated total energy using the third-order Birch-Murnaghan
equation of states \cite{eos} separately for the low- and high-volume regions.
Overall, our results for the electronic structure and lattice properties of FeO,
which now include the effect of electronic correlations, agree well with
experimental data. 
In particular, our calculations at ambient pressure give a Mott insulating
solution with an energy gap of $\sim 0.8$ eV. The energy gap lies between the
top of the valence band originating from the mixed Fe $3d$ and O $2p$ states and
empty Fe $4s$ states. Our result for the Mott $d$-$d$ energy gap is about 2 eV,
in good agreement with optical and photoemission experiments. \cite{Spectra}
We find the equilibrium lattice constant $a = 8.36$ a.u., which is less than 1-2~\% 
off the experimental value. The calculated bulk modulus is $B = $ 140 GPa,
the local magnetic moment $\sqrt{\langle m_z^2 \rangle} \sim
3.7~\mu_B$. Fe $t_{2g}$ and $e_g$ orbital occupancies are 0.68 and 0.55,
respectively. These findings clearly indicate that at ambient pressure Fe$^{2+}$
ion is in a high-spin state $(S=2)$. In fact, in a cubic crystal field, the
Fe$^{2+}$ ions (i.e., Fe $3d^6$ configuration with four electrons in the
$t_{2g}$ and two in the $e_g$ orbitals) have a local moment of 4~$\mu_B$.

Furthermore, the total-energy and local-moment calculation results exhibit a
remarkable anomaly upon compression of the lattice volume down to $\nu \sim
0.72$.
Indeed, the local moment is seen to retain its high-spin value down to about 
$\nu \sim 0.8$. Upon further compression, a broad high-spin (HS) to low-spin 
(LS) crossover takes place, with a collapse of the local moment to a LS state 
with magnetic moment $\sim 1.2~\mu_B$ at pressure above 160~GPa, i.e.,
$\nu < 0.6$.
We display our results for the evolution of the Fe $t_{2g}$ and $e_g$ orbital
occupations in Fig.~\ref{Fig_1} (bottom). Upon compression, we observe a 
substantial redistribution
of charge between the $t_{2g}$ and $e_g$ orbitals within the Fe $3d$ shell. Fe
$t_{2g}$ orbital occupations are found to gradually increase with pressure,
resulting in (almost) completely occupied state ($t_{2g}$ occupation is about
0.95). On the other hand, the $e_{g}$ orbitals are strongly depopulated (their
occupation is 0.25) and the Fe $3d$ total occupancy remains essentially unchanged
with pressure. We therefore interpret this spin crossover as a HS-LS
transition. 

The spin-state transition is accompanied by a structural transformation. The structural 
change takes place upon a compression of the lattice volume to $\nu \sim 0.72$, 
resulting in a collapse of the lattice volume by $\sim 8.5$ \%. 
This value should be considered as an upper-bound estimate because we
neglect multiple intermediate-phase transitions when fit the total-energy result
to the third-order Birch-Murnaghan equation of states.
Our estimate for the transition pressure is $p = 73$ GPa. 
In addition, we note that the bulk modulus in the HS phase is somewhat smaller
than that in the LS phase (162~GPa), implying an enhancement of the
compressibility at the phase transition. 
Overall, the electronic structure, the equilibrium lattice constant, and the
structural phase stability of paramagnetic FeO obtained by the fully charge
self-consistent GGA+DMFT approach are in remarkably good agreement with the
experimental data. Our findings clearly indicate the crucial importance of
electronic correlations to explain the electronic structure and lattice
properties of paramagnetic FeO.


\begin{figure}[tbp!]
\centerline{\includegraphics[width=0.5\textwidth,clip=true]{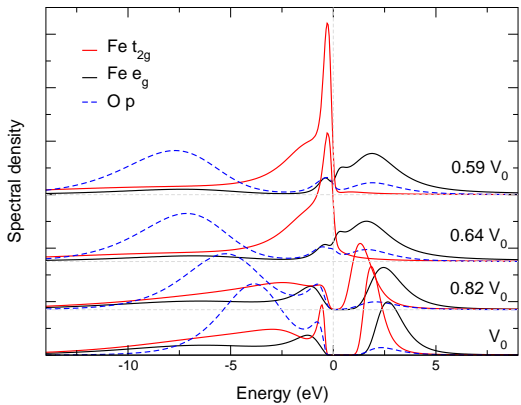}}
\caption{Evolution of the spectral function of paramagnetic FeO as a function of
lattice volume. Fe $t_{2g}$/$e_g$ and O $p$ orbital contributions are shown. The
metal-insulator transition associated with the HS-LS transformation takes place
at $\nu \sim 0.72$, at pressure $\sim 73$ GPa.}
\label{Fig_2}
\end{figure}

Next we address the spectral properties of paramagnetic FeO. In Fig.~\ref{Fig_2}
we present the evolution of the spectral function of FeO as a function of
lattice volume.
We note that at ambient pressure paramagnetic FeO is a Mott insulator with a
relatively large $d$-$d$ gap $\sim 2$ eV. Interestingly, the energy gap lies
between the Fe $t_{2g}$ states, while the $t_{2g}$-$e_g$ gap is somewhat larger,
$\sim 2.5$~eV.
The top of the valence band has a mixed Fe $3d$ and O $2p$ character, with a
resonant peak in the filled $t_{2g}$ band located at about -0.9 eV. The latter can
be ascribed to the formation of a Zhang-Rice bound state. \cite{ZR88}
We find the $d$-$d$ energy gap being gradually decreased upon compression,
resulting in an orbital-selective Mott metal-insulator transition (MIT). In
fact, the onset of the local moment collapse at $\nu \sim 0.75$ is associated
with closing of the gap for the $t_{2g}$ orbitals, while the $e_g$ states remain
to be gaped down to $\nu \sim 0.7$.
We note that the HS-LS transition takes place upon compression of the lattice
volume by $\nu \sim 0.72$, at critical pressure of $p \sim 73$ GPa, in agreement
with experiment.
Our calculations clearly show that the HS-LS state transition in the B1
structure of paramagnetic FeO is associated with an orbital-selective
Mott-Hubbard MIT. Moreover, the MIT is accompanied by a collapse of the lattice
volume by about $8.5$~\%.


In addition, we check how the electronic structure of FeO depend on a
(possible) reduction of the Coulomb interaction $U$ with pressure. For this
purpose, we perform calculations with a substantially smaller value $U=5$~eV 
and the same Hund's coupling value $J=0.89$~eV. Our new results are in overall 
qualitative
agreement with those presented above. We notice that the HS-LS crossover is
accompanied by an orbital-selective MIT which is found to occur at somewhat
smaller compression $\nu \sim 0.77$, at critical pressure 55~GPa. We also note
a substantial reduction of the Mott $d$-$d$ band gap to about 1.1~eV, whereas
the charge-transfer gap between the Fe $3d$-O $2p$ and empty Fe $4s$ states
remains unchanged, about 0.8~eV. These findings clearly indicate that our
results for the spin-state crossover and associated with it the
orbital-selective MIT are robust. The HS to LS transition takes place even at
a substantially reduced (possibly due to the applied pressure) value of $U$.


\begin{figure}[tbp!]
\centerline{\includegraphics[width=0.5\textwidth,clip=true]{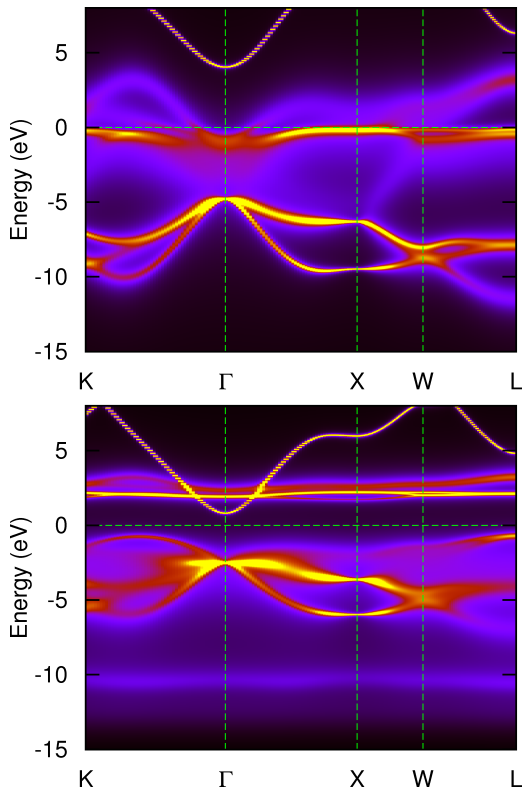}}
\caption{The {\bf k}-resolved spectral function of paramagnetic FeO computed
within GGA+DMFT for different volumes. Top: Results for the LS phase with
lattice volume $\nu \sim 0.6$. Bottom: The equilibrium HS phase.}
\label{Fig_3}
\end{figure}

We also calculated the momentum-resolved spectral function of paramagnetic FeO.
In Fig.~\ref{Fig_3} we present our results obtained for the B1 structure of
paramagnetic FeO across the MIT. Our calculations at ambient pressure show a
Mott insulator with an energy gap value of $\sim 0.8$~eV [see Fig.~\ref{Fig_3}
(bottom)]. The energy gap lies between the occupied states with a mixed Fe
$3d$-O $p$ character and empty Fe $4s$ state (the latter is clearly seen as a
broad parabolic-like band at the $\Gamma$-point just above the Fermi energy), in
agreement with photoemission and optical experiments. \cite{Spectra}
In fact, the optical spectroscopy measurements observe a weak absorption between
0.5 and 2.0 eV, assigned to the Fe $3d$-O $2p$ to Fe $4s$ transitions. The strong
absorption edge associated with the $d$-$d$ transitions is found to appear in
optical spectroscopy at about 2.4 eV. 
Our result for the $d$-$d$ gap at ambient pressure is about 2 and 2.5~eV for the
$t_{2g}$-$t_{2g}$ and $t_{2g}$-$e_{g}$ transitions, respectively.
The O $2p$ states are about -4~eV below the Fermi level, but still have a
substantial contribution near the Fermi level. We note however that the Fe
$t_{2g}$ states are seen to have a predominant contribution around the Fermi
level. We therefore interpret paramagnetic FeO as a Mott-Hubbard insulator, in
agreement with previous studies.
Furthermore, our calculations show an entire reconstruction of the electronic
structure of paramagnetic FeO in the LS phase [see Fig.~\ref{Fig_3} (top)].
Namely, we obtain a strongly correlated metal with almost completely occupied
$t_{2g}$ band, which is located in the vicinity of the Fermi level, and a broad
disperse spectral weight originating from the Fe $e_g$ band crossing the Fermi
level.
Moreover, the Fe $t_{2g}$ states exhibit a resonant state just below the Fermi
level which can be ascribed to the Zhang-Rice bound state.
The O $2p$ states are shifted to about -8~eV below the Fermi level, and now have
a moderate contribution near the Fermi level.

\begin{figure}[tbp!]
\centerline{\includegraphics[width=0.5\textwidth,clip=true]{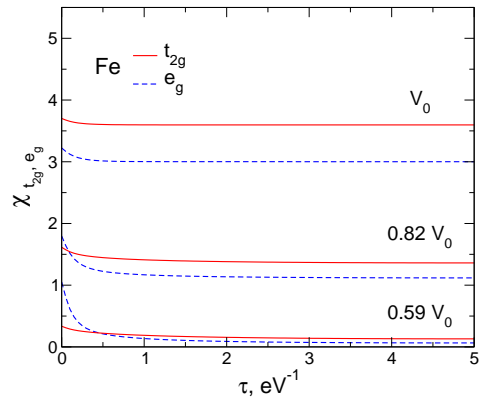}}
\caption{Local spin-spin correlation function $\chi(\tau)$ calculated by
GGA+DMFT for paramagnetic FeO as a function of volume. The intra-orbital
$t_{2g}$ and $e_g$ contributions are shown.
}
\label{Fig_4}
\end{figure}

We also calculate the local (dynamical) spin-spin correlation function
$\chi(\tau)=\langle \hat m_z(\tau)\hat m_z(0) \rangle$ of paramagnetic 
FeO for different volumes, where $\tau$ is the imaginary time. 
In Fig.~\ref{Fig_4} we display our results 
for the corresponding intra-orbital $t_{2g}$ and $e_g$ contributions. 
We note that the $t_{2g}$ contributions are seen to be almost independent 
of $\tau$, even in the LS state, implying that the $t_{2g}$ moment remains
localized across the HS-LS transition. 
On the other hand, the $e_g$ states clearly exhibit a crossover from  localized
to itinerant magnetic behavior under pressure, suggesting that the HS-LS
metal-insulator transition in paramagnetic FeO is of the orbital-selective type.
In addition, we notice that at high pressures the screened local moment defined
as $\propto \int_0^{1/T} d\tau\chi(\tau)$ differs from the corresponding
instantaneous ${\langle m_z^2 \rangle}$ value (mostly because of the
contribution originating from the $e_g$ band), which suggests an
itinerant-moment behavior of the LS state. 


\begin{figure}[tbp!]
\centerline{\includegraphics[width=0.5\textwidth,clip=true]{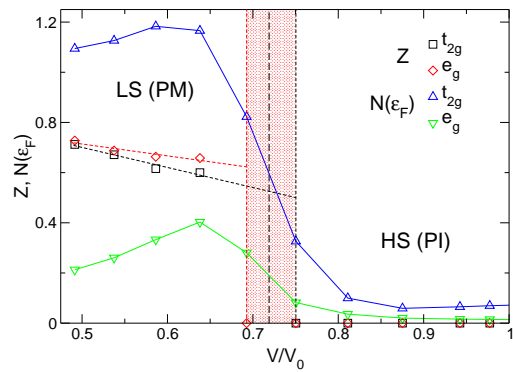}}
\caption{Quasiparticle weight $Z$ and spectral weight at the Fermi level
$N(\epsilon_F) = -\frac{\beta}{\pi} G(\tau=\beta/2)$ calculated for the Fe
$t_{2g}$ and $e_g$ orbitals across the MIT in paramagnetic FeO. The structural
change associated with the HS-LS transition is indicated by a vertical black
dashed line. The orbital-selective MIT phase is marked by a red filled
rectangle.}
\label{Fig_5}
\end{figure}

Finally, we calculate the quasiparticle weight employing a polynomial fit of the
imaginary part of the self-energy $\Sigma(i\omega_n)$ at the lowest Matsubara
frequencies $\omega_n$. It is evaluated as $Z=[1 - \partial Im
\Sigma(i\omega)/\partial i\omega]^{-1}$ from the slope of the polynomial fit at
$\omega=0$. In Fig.~\ref{Fig_5} we present our results for the partial $t_{2g}$
and $e_g$ contributions computed for different volumes. In the LS phase, the
$t_{2g}$ and $e_g$ quasiparticle weights are finite and found to decrease
monotonously upon expansion of the lattice.  
Upon further expansion of the lattice, the $t_{2g}$ and $e_g$ $Z$ factors are
seen to collapse to zero at different volumes. This result again suggests that the
HS-LS transition is accompanied by an orbital-selective MIT. The electronic
effective mass diverges at the MIT, in accordance with a Brinkman-Rice picture
of the MIT. \cite{BR71}
We note that this divergence coincides with the drop of the spectral weight for
the $t_{2g}$ and $e_g$ orbitals at the Fermi level shown in Fig.~\ref{Fig_5}.

\section{Conclusion}

We determined the electronic structure and phase stability of the
B1 crystal structure of paramagnetic FeO at high pressure and temperature. Our
results clearly establish a high-spin to low-spin transition which is found to
appear upon compression of the lattice by $V/V_0 \sim 0.72$, at pressure 73~GPa,
in agreement with experiment. The spin-state transition is intimately linked
with an orbital-selective Mott-Hubbard metal-insulator transition. 
The MIT is accompanied by a collapse of the lattice volume by $\sim 8.5$~\%, 
implying a complex interplay between electronic and lattice degrees of freedom.
Upon compression, the Fe $e_g$ states exhibit a crossover from a localized 
to itinerant-moment behavior, while the $t_{2g}$ moment remains localized 
across the HS-LS transition.
Our results are consistent with the picture of an electronically driven
spin-state transition and volume collapse in the B1 crystal structure of
paramagnetic FeO.

\begin{acknowledgments}

We thank V. I. Anisimov, and D. Vollhardt for valuable discussions. Support by
the Deutsche Forschungsgemeinschaft through Transregio TRR 80 is gratefully acknowledged.

\end{acknowledgments}


\begin{thebibliography}{99}


\bibitem{Phase-diagram} 
H. Mao, J. Shu, Y. Fei, J. Hu, and R. J. Hemley, Phys. Earth Planet. Inter. \textbf{96}, 135 (1996);
R. A. Fischer. A. J. Campbell. O. T. Lord, G. A. Shofner. P.
Dera, and V. B. Prakapenka, Geophys. Res. Lett. \textbf{38}, L24301 (2011);
R. A. Fischer, A. J. Cambell. G. A. Shofner, O. T. Lord, P. Dera, V. P. Prakapenka, Earth Planet. Sci. Lett. \textbf{304}, 496 (2011);
K. Ohta, K. Fujino, Y. Kuwayama, T. Kondo, K. Shimizu, and Y.
Ohishi, J. Geophys. Res. Solid Earth, \textbf{119}, doi:10.1002/2014JB01.0972.

\bibitem{T-Neel} 
C. A. McCammon, J. Magn. Magn. Mater. \textbf{104}, 1937 (1992);
A. P. Kantor, S. D. Jacobsen, I. Yu. Kantor, L. S. Dubrovinsky,
C. A. McCammon, H. J. Reichmann, and I. N. Goncharenko, Phys. Rev. Lett.
\textbf{93}, 215502 (2004).


\bibitem{rB1-phase} 
B. T. M. Willis and H. P. Rooksby, Acta Crystallogr. \textbf{6}, 827 (1953).
C. A. McCammon and L. Liu, Phys. Chem. Miner. \textbf{10}, 106 (1984);
T. Yagi, T. Suzuki, and S. Akimoto, J. Geophys. Res. \textbf{90}, 8784 (1985).
S. Ono, Y. Ohishi, and T. Kikegawa, J. Phys. Condens. Matter \textbf{19}, 036205 (2007).

\bibitem{B8-phase}
Y. Fei and H.-K. Mao, Science \textbf{266}, 1678 (1994);
K. Ohta, K. Hirose, K. Shimizu, and Y. Ohishi, Phys. Rev. B \textbf{82}, 174120 (2010);
H. Ozawa, K. Hirose, S. Tateno, N. Sata, and Y. Ohishi, Phys. Earth Planet. Inter. \textbf{179} 157 (2010);
\bibitem{OH11} H. Ozawa, K. Hirose, K. Ohta, H. Ishii, N. Hiraoka, Y. Ohishi,
and Y. Seto, Phys. Rev. B \textbf{84}, 134417 (2011).


\bibitem{IC93} 
D. G. Isaak, R. E. Cohen, M. J. Mehl, and D. J. Singh, Phys. Rev. B \textbf{47}, 7720 (1993).
R. E. Cohen, I. I. Mazin, and D. G. Isaak,  Science \textbf{275}, 654 (1997).

\bibitem{MA97} 
I. I. Mazin and V. I. Anisimov, Phys. Rev. B \textbf{55}, 12822 (1997).

\bibitem{MF98} I. I. Mazin, Y. Fei, R. Downs, and R. Cohen, Amer. Mineral.
\textbf{83}, 451 (1998).

\bibitem{ZT98} Z. Fang, K. Terakura, H. Sawada, T. Miyazaki, and I. Solovyev,
Phys. Rev. Lett. \textbf{81}, 1027 (1998);
Z. Fang, I. V. Solovyev, H. Sawada, and K. Terakura, Phys. Rev. B \textbf{59},
762 (1999).

\bibitem{GC03} S. A. Gramsch, R. E. Cohen, and S. Yu. Savrasov, Amer. Mineral.
\textbf{88}, 257 (2003).

\bibitem{TB06} F. Tran, P. Blaha, K. Schwarz, and P. Nov\'ak, Phys. Rev. B \textbf{74}, 155108 (2006).

\bibitem{KM08} J. Kolorenc and L. Mitas, Phys. Rev. Lett. \textbf{101}, 185502
(2008).

\bibitem{SP10} A. O. Shorikov, Z. V. Pchelkina, V. I. Anisimov, S. L.
Skornyakov, and M. A. Korotin, Phys. Rev. B \textbf{82}, 195101 (2010).

\bibitem{OC12} K. Ohta, R. E. Cohen, K. Hirose, K. Haule, K. Shimizu, and Y.
Ohishi, Phys. Rev. Lett. \textbf{108}, 026403 (2012).

\bibitem{HS15} E. Holmstr\"om and L. Stixrude, Phys. Rev. Lett. \textbf{114},
117202 (2015).


\bibitem{KJ86} E. Knittle, R. Jeanloz, A. C. Mitchell, and W. J. Nellis, Solid
State Commun. \textbf{59}, 513 (1986).

\bibitem{PT97} M. P. Pasternak, R. D. Taylor, R. Jeanloz, X. Li, J. H. Nguyen,
and C. A. McCammon, Phys. Rev. Lett. \textbf{79}, 5046 (1997).

\bibitem{BS99} J. Badro, V. V. Struzhkin, J. Shu, R. J. Hemley, H.-k. Mao, C.-C.
Kao, J.-P. Rueff, and G. Shen, Phys. Rev. Lett. \textbf{83}, 4101 (1999)

\bibitem{MR07} A. Mattila, J.-P. Rueff, J. Badro, G. Vank\'o, and A. Shukla,
Phys. Rev. Lett. \textbf{98}, 196404 (2007).

\bibitem{FC10}
R. A. Fischer and A. J. Campbell, Am. Mineral. \textbf{95}, 1473 (2010).

\bibitem{DMFT} W. Metzner and D. Vollhardt, Phys. Rev. Lett. \textbf{62}, 324 (1989);
A. Georges, G. Kotliar, W. Krauth, and M. J. Rozenberg, Rev. Mod. Phys. \textbf{68}, 13 (1996);
G. Kotliar and D. Vollhardt, Phys. Today \textbf{57}(3), 53 (2004).

\bibitem{LDA+DMFT} V. I. Anisimov, A. I. Poteryaev, M. A. Korotin, A. O.
Anokhin, and G. Kotliar, J. Phys. Condens. Matter \textbf{9}, 7359 (1997); 
A. I. Lichtenstein and M. I. Katsnelson, Phys. Rev. B \textbf{57}, 6884 (1998);
G. Kotliar, S. Y. Savrasov, K. Haule, V. S. Oudovenko, O. Parcollet, and C. A. 
Marianetti, Rev. Mod. Phys. \textbf{78}, 865 (2006).

%
%
\bibitem{SK01}
S. Y. Savrasov, G. Kotliar, and E. Abrahams, Nature \textbf{410}, 793 (2001);
X. Dai, S. Y. Savrasov, G. Kotliar, A. Migliori, H. Ledbetter, and E. Abrahams, Science \textbf{9}, 953, (2003).

\bibitem{HK04}
K. Held, G. Keller, V. Eyert, D. Vollhardt, and V. I. Anisimov, Phys. Rev. Lett. \textbf{86}, 5345 (2001); 
G. Keller, K. Held, V. Eyert, D. Vollhardt, and V. I. Anisimov, Phys. Rev. B \textbf{70}, 205116 (2004).

\bibitem{HM01}
K. Held, A. K. McMahan, and R. T. Scalettar, Phys. Rev. Lett. \textbf{87}, 276404 (2001);
A. K. McMahan, K. Held, and R. T. Scalettar, Phys. Rev. B \textbf{67}, 075108 (2003).

\bibitem{SH06}
S. Y. Savrasov, K. Haule, and G. Kotliar, Phys. Rev. Lett. \textbf{96}, 036404 (2006).

\bibitem{AB06}
B. Amadon, S. Biermann, A. Georges, and F. Aryasetiawan, Phys. Rev. Lett. \textbf{96}, 066402 (2006);
N. Lanat\'a, Y.-X. Yao, C.-Z. Wang, K.-M. Ho, J. Schmalian, K. Haule, and G. Kotliar,
Phys. Rev. Lett. \textbf{111}, 196801 (2013);
J. Bieder and B. Amadon, Phys. Rev. B \textbf{89}, 195132 (2014);
B. Chakrabarti, M. E. Pezzoli, G. Sordi, K. Haule, and G. Kotliar,
Phys. Rev. B \textbf{89}, 125113 (2014).

\bibitem{KA07}
J. Kunes, V. I. Anisimov, A. V. Lukoyanov, and D. Vollhardt, Phys. Rev. B \textbf{75}, 165115 (2007);
J. Kunes, V. I. Anisimov, S. L. Skornyakov, A. V. Lukoyanov, and D. Vollhardt,
Phys. Rev. Lett. \textbf{99}, 156404 (2007). 

\bibitem{LB08}
I. Leonov, N. Binggeli, D. Korotin, V. I. Anisimov, N. Stoji\'c, and D. Vollhardt, Phys. Rev. Lett. \textbf{101}, 096405 (2008);
I. Leonov, Dm. Korotin, N. Binggeli, V. I. Anisimov, and D. Vollhardt, Phys. Rev. B \textbf{81}, 075109 (2010);
J. Kunes, I. Leonov, M. Kollar, K. Byczuk, V. I. Anisimov, D. Vollhardt, Eur. Phys. J. Spec. Top. \textbf{180}, 5 (2010).

\bibitem{KL08}
J. Kunes, A. V. Lukoyanov, V. I. Anisimov, R. T. Scalettar, and W. E. Pickett, Nat. Mater. \textbf{7}, 198 (2008);
J. Kunes, Dm. M. Korotin, M. A. Korotin, V. I. Anisimov, and P. Werner, Phys. Rev. Lett. \textbf{102}, 146402 (2009).

\bibitem{DM09}
I. Di Marco, J. Min\'ar, S. Chadov, M. I. Katsnelson, H. Ebert, and A. I. Lichtenstein,
Phys. Rev. B \textbf{79}, 115111 (2009).

\bibitem{LP11}
I. Leonov, A. I. Poteryaev, V. I. Anisimov, and D. Vollhardt, Phys. Rev. Lett. \textbf{106}, 106405 (2011);
I. Leonov, A. I. Poteryaev, V. I. Anisimov, and D. Vollhardt, Phys. Rev. B \textbf{85}, 020401(R) (2012);
I. Leonov, A. I. Poteryaev, Yu. N. Gornostyrev, A. I. Lichtenstein, M. I. Katsnelson, V. I. Anisimov, D. Vollhardt, Sci. Rep. \textbf{4}, 5585 (2014).

\bibitem{AP11}
M. Aichhorn, L. Pourovskii, and A. Georges, Phys. Rev. B \textbf{84}, 054529 (2011).

\bibitem{HW12}
L. Huang, Y. Wang, and X. Dai, Phys. Rev. B \textbf{85}, 245110 (2012);
A. A. Dyachenko, A. O. Shorikov, A. V. Lukoyanov, and V. I. Anisimov, JETP Lett. \textbf{96}, 56 (2012).

\bibitem{GP12}
D. Grieger, C. Piefke, O. E. Peil, and F. Lechermann, Phys. Rev. B \textbf{86}, 155121 (2012);
D. Grieger and F. Lechermann, \emph{ibid.} \textbf{90}, 115115 (2014);
I. Leonov, V. I. Anisimov, and D. Vollhardt, Phys. Rev. B \textbf{91}, 195115 (2015). 

\bibitem{PMM14}
H. Park, A. J. Millis, C. A. Marianetti, Phys. Rev. B \textbf{89}, 245133 (2014).

\bibitem{PM14}
K. Glazyrin, L. V. Pourovskii, L. Dubrovinsky, O. Narygina, C. McCammon, B. Hewener, V. Sch\"unemann, J. Wolny, K. Muffler, 
A. I. Chumakov, W. Crichton, M. Hanfland, V. B. Prakapenka, F. Tasnadi, M. Ekholm, M. Aichhorn, V. Vildosola, A. V. Ruban, 
M. I. Katsnelson, and I. A. Abrikosov, Phys. Rev. Lett. \textbf{110}, 117206 (2013);
L. V. Pourovskii, J. Mravlje, M. Ferrero, O. Parcollet, and I. A. Abrikosov, Phys. Rev. B \textbf{90}, 155120 (2014).

\bibitem{LA14}
I. Leonov, V. I. Anisimov, and D. Vollhardt, Phys. Rev. Lett. \textbf{112}, 146401 (2014).


\bibitem{charge-sc-LDA+DMFT} For a review, see, e.g., L. V. Pourovskii, B.
Amadon, S. Biermann, and A. Georges, Phys. Rev. B \textbf{76}, 235101 (2007); K.
Haule, \emph{ibid.} \textbf{75}, 155113 (2007); B. Amadon, F. Lechermann, A.
Georges, F. Jollet, T. O. Wehling, and A. I. Lichtenstein, \emph{ibid.}
\textbf{77}, 205112 (2008); M. Aichhorn, L. Pourovskii, V. Vildosola, M.
Ferrero, O. Parcollet, T. Miyake, A. Georges, and S. Biermann, \emph{ibid.}
\textbf{80}, 085101 (2009); B. Amadon, J. Phys. Condens. Matt. \textbf{24}, 
075604 (2012); H. Park, A. J. Millis, and C. A. Marianetti, Phys. Rev. B 
\textbf{90}, 235103 (2014). 

\bibitem{Pseudo}
S. Baroni, S. de Gironcoli, A. Dal Corso, and P. Giannozzi,
Rev. Mod. Phys. \textbf{73}, 515 (2001); P. Giannozzi, S. Baroni, N. Bonini, M.
Calandra, R. Car \emph{et al.}, J. Phys. Condens. Matter \textbf{21}, 395502
(2009).

\bibitem{MV97} 
N. Marzari and D. Vanderbilt, Phys. Rev. B \textbf{56}, 12847 (1997);
N. Marzari, A. A. Mostofi, J. R. Yates, I. Souza, and D.
Vanderbilt, Rev. Mod. Phys. \textbf{84}, 1419 (2012).

\bibitem{Wannier-functions} 
V. I. Anisimov, D. E. Kondakov, A. V. Kozhevnikov, I. A. Nekrasov, Z. V.
Pchelkina, J. W. Allen, S.-K. Mo, H.-D. Kim, P. Metcalf, S. Suga, A. Sekiyama,
G. Keller, I. Leonov, X. Ren, and D. Vollhardt, Phys. Rev. B \textbf{71}, 125119
(2005);
G. Trimarchi, I. Leonov, N. Binggeli, Dm. Korotin, and V. I. Anisimov, J. Phys.
Condens. Matter \textbf{20}, 135227 (2008); Dm. Korotin, A. V. Kozhevnikov, S.
L. Skornyakov, I. Leonov, N. Binggeli, V. I. Anisimov and G. Trimarchi, Eur.
Phys. J. B \textbf{65}, 91 (2008).

\bibitem{ctqmc}
P. Werner, A. Comanac, L. de'Medici, M. Troyer, and A. J. Millis, Phys. Rev.
Lett. \textbf{97}, 076405 (2006); E. Gull, A. J. Millis, A. I. Lichtenstein, A.
N. Rubtsov, M. Troyer, and P. Werner, Rev. Mod. Phys. \textbf{83}, 349 (2011).

\bibitem{Spectra}
P. S. Bagus, C. R. Brundle, T. J. Chuang, and K. Wandelt
Phys. Rev. Lett. \textbf{39}, 1229 (1977);
I. Balberg and H. L. Pinch, J. Magn. Magn. Mater. \textbf{7}, 12 (1978).

\bibitem{eos}
F. D. Murnaghan, Proc. Natl. Acad. Sci. U.S.A. \textbf{30}, 244 (1944);
F. Birch, Phys. Rev. \textbf{71}, 809 (1947).

\bibitem{ZR88} F. C. Zhang and T. M. Rice, Phys. Rev. B \textbf{37}, 3759(R) (1988);
J. Bala, A. M. Ole\'s, and J. Zaanen, Phys. Rev. Lett. \textbf{72}, 2600 (1994);
Q. Yin, A. Gordienko, X. Wan, and S. Y. Savrasov, \emph{ibid.} \textbf{100}, 066406 (2008).

\bibitem{BR71} W. F. Brinkman and T. M. Rice, Phys. Rev. B \textbf{2}, 4302 (1970).


 
\end{thebibliography}
\end{document}